\documentclass[twocolumn,trackchanges]{aastex631}

\usepackage{amsmath,xcolor}
\usepackage{amsmath} 
\usepackage{array}
\usepackage{bm}

\usepackage{amsmath} 

\begin{document}

\title{Irradiated Atmospheres II: Interplay Between 
Scattering \\ 
and Vertical-Mixing induced Energy Transport
}

\correspondingauthor{Cong Yu}
\email{yucong@mail.sysu.edu.cn}

\author[0009-0004-1986-2185]{Zhen-Tai Zhang}
\affiliation{School of Physics and Astronomy, Sun Yat-Sen University, Zhuhai, 519082, People's Republic of China}
\affiliation{CSST Science Center for the Guangdong-Hong Kong-Macau Greater Bay Area, Zhuhai, 519082, People's Republic of China}
\affiliation{State Key Laboratory of Lunar and Planetary Sciences, Macau University of Science and Technology, Macau, People's Republic of China}

\author[0000-0002-0447-7207]{Wei Zhong}
\affiliation{School of Physics and Astronomy, Sun Yat-Sen University, Zhuhai, 519082, People's Republic of China}
\affiliation{CSST Science Center for the Guangdong-Hong Kong-Macau Greater Bay Area, Zhuhai, 519082, People's Republic of China}
\affiliation{State Key Laboratory of Lunar and Planetary Sciences, Macau University of Science and Technology, Macau, People's Republic of China}

\author[0000-0003-2278-6932]{Xianyu Tan}
\affiliation{Tsung-Dao Lee Institute \& School of Physics and Astronomy, Shanghai Jiao Tong University, Shanghai 201210, China}

 \author[0000-0002-0378-2023]{Bo Ma}
 \affiliation{School of Physics and Astronomy, Sun Yat-Sen University, Zhuhai, 519082, People's Republic of China}
 \affiliation{CSST Science Center for the Guangdong-Hong Kong-Macau Greater Bay Area, Zhuhai, 519082, People's Republic of China}

\author{Ruyi Wei}
 \affiliation{Electric Information School, Wuhan University, Wuhan, 430072, People's Republic of China}

\author[0000-0003-0454-7890]{Cong Yu}
\affiliation{School of Physics and Astronomy, Sun Yat-Sen University, Zhuhai, 519082, People's Republic of China}
\affiliation{CSST Science Center for the Guangdong-Hong Kong-Macau Greater Bay Area, Zhuhai, 519082, People's Republic of China}
\affiliation{State Key Laboratory of Lunar and Planetary Sciences, Macau University of Science and Technology, Macau, People's Republic of China}
\affiliation{International Centre of Supernovae, Yunnan Key Laboratory, Kunming 650216, People's Republic of China}
\begin{abstract}
The scattering is crucial for the atmospheric thermal profiles.
The energy transport by the vertical mixing plays an essential role for the greenhouse or anti-greenhouse effect. 
This work explores the interaction between scattering and vertical mixing, specifically whether these processes enhance or mitigate each other's effects on atmospheric temperature. 
The interaction between mixing flux and scattering is nonlinear. 
Our calculations indicate that thermal scattering intensifies the greenhouse effects caused by vertical mixing in the middle atmosphere but reduces it in the lower layers.
In the middle atmosphere, increased vertical mixing enhances the warming effect of the thermal scattering while diminishing the cooling effect of visible scattering. 
In the lower atmosphere, it enhances the anti-greenhouse effect linked to visible scattering and diminishes the greenhouse effect produced by thermal scattering. 
The combined influence of thermal scattering and vertical mixing on the lower atmosphere's greenhouse effect is weaker than their separate impacts, akin to $1+1<2$.
It is also interesting to note that the joint effect may also influence chemistry and cloud formation, altering the thermal structure.
\end{abstract}


\section{Introduction}
Since the groundbreaking identification of 51 Pegasi b \citep{1995Mayor}, over 5,700 exoplanets have been found.
Characterization of these planetary atmospheres is predominantly achieved through transmission and emission photometry \citep{2010Seager}.
With the deployment of the James Webb Space Telescope (JWST, \citealt{2023Carter,2023Miles}), the precision of detecting exoplanetary atmospheres has improved substantially.
These high-precision observations facilitate thorough analysis of atmospheric characteristics, with particular emphasis on clouds, which are commonly found in exoplanetary atmospheres \citep{2016Iyer,2015Wakeford,2023Taylor}. 
Their scattering properties are crucial in determining the spectral features of exoplanets \citep{2011deKok,2020Molli,2021Taylor,2023Singla}.\par
Clouds and hazes play a vital role in the analysis of the combined transit spectrum from the JWST and HST observations of WASP-39b \citep{2023Arfaux}. 
In fact, these dominate the scattering properties.
\cite{2023Taylor} demonstrated that a scattered-cloud model more accurately accounts for the day-side emission spectrum of WASP-43b compared to a cloud-free model.
Moreover, scattering-induced diffuse transmission significantly affects transmission spectra \citep{2023Singla}. In parallel, \cite{2021Taylor} suggested that scattering generates distinctive features in emission spectra, even within isothermal atmospheres.
Their JWST simulations identified that retrieval degeneracies and biases arise when cloud scattering dominates the spectral profile. 
Furthermore, continuum scattering has a substantial impact on both the radiation field and the thermal profile, leading to notable shifts in atmospheric temperatures \citep{Heng2014, 2018Mohandas}. Specifically, visible scattering generally decreases the thermal state of lower atmospheric layers and causes the anti-greenhouse effect, whereas thermal scattering elevates atmospheric temperatures, thus amplifying the greenhouse effect. \par

In the atmosphere, atoms and molecules possess sizes that are significantly smaller than the wavelength of general radiation. In such instances, the scattering phenomena are characterized as Rayleigh scattering, with the scattering cross section for each particle being proportional to $\lambda^{-4}$. Therefore, visible radiation with shorter wavelengths is subject to stronger scattering. 
Larger particles are required in the atmosphere to make scattering of longer wavelengths significant. However, for larger particles such as aerosols, haze, and clouds, the scattering is no longer merely Rayleigh scattering. Alternative models are required to address this complexity, such as the Mie scattering theory \citep{1908Mie}. Moreover, due to the irregular shapes of these particles, many of their scattering properties can only be determined experimentally. Generally, when aerosols, haze, and clouds are present in the atmosphere, they contribute to both visible and thermal scattering.

Dynamic mixing is a key factor in shaping atmospheric chemistry and cloud formation.
The variations in chemistry and cloud from the day to night sides of a hot Jupiter's atmosphere arise from varying levels of received irradiation. These differences are modified by horizontal mixing \citep{2024Powell}, a factor typically addressed in models that go beyond 1D frameworks. 
 In this study, we employ a 1D radiative transport model to investigate vertical mixing.
Vertical mixing shapes the chemistry and clouds of the atmosphere by transporting material vertically \citep{2018Gao, 2001Ackerman, 2007Hubeny}.

Dynamic mixing also contributes to energy transport within the atmosphere. Horizontal mixing is responsible for the day-to-night energy flux \citep{2021Parmentier} in low-pressure regions, while vertical mixing generates a vertical energy flux, which is directed downward in the radiative region \citep{Youdin2010,2018Leconte}. This downward flux triggers the ``mechanism greenhouse effect" and increases the apparent radius of the planet \citep{2017Tremblin,2019Sainsbury,2021Fortney}. 
In the radiative layer, vertical mixing is influenced by atmospheric circulation \citep{1984maph...Holton} and breaking gravity waves \citep{1987JGR....Strobel}. 
The mixing flux shifts the radiative-convective boundary to higher pressures, compressing the convective layer deeper within the atmosphere \citep{Youdin2010}. 
Moreover, the mixing flux interacts with other physical processes, further shaping the atmospheric thermal profile.\par
 In Paper I \citep{2024Zhong}, we show the impact of vertical-mixing-induced energy transport on the temperature.
 , which had not been explored before.
The individual impacts of each are relatively straightforward.
However, the effects of scattering and vertical mixing energy transport are intertwined.
Scattering alters the magnitude of the mixing flux, which is determined by the logarithmic temperature gradient and density \citep{Youdin2010}. 
Consequently, temperature adjustments from vertical mixing vary with scattering conditions, affecting the temperature-shift ratio.
The intensities of the mixing flux and scattering are systematically varied in our calculations to reveal their mutual effects.\par
Non-isotropic scattering complicates the solution of the radiative transfer equation, as it often requires expanding many terms in the scattering phase function, leading to a large system of equations \citep{stamnes2017}.
To address these computational difficulties, various approximation techniques have been introduced, including the $\delta-TTA$ scaling transformation, which is also known as the two-term Delta-Eddington approximation \citep{1976Joseph}. 
The $TTA$ method approximates only the first two terms of the scattering phase function and is commonly used alongside the two-stream approximation\citep{Pierrehumbert2010, Heng2014, Heng2017}, significantly simplifying calculations. 
In this work, the hemispheric two-stream solution\citep{2008Hansen, 2010Guillot, Heng2014, Heng2017} is applied to address non-isotropic scattering, and the semi-grey model is adopted to calculate radiative transport. 
To quantify the mixing flux, the formulation proposed by \cite{Youdin2010} is adopted.\par
The framework of this work is as follows:
\S \ref{method} presents the models and equations utilized in this study; 
\S \ref{result} details the analytical formulas and numerical solutions, describing the atmospheric temperature incorporating both vertical mixing and non-isotropic scattering under the two-stream approximation.
Finally, \S \ref{conclusion} provides a comprehensive summary and discussion of the findings.

\section{Atmosphere with vertical mixing and coherent scattering}
\label{method}

This section shows the radiative equations relevant to a plane-parallel atmosphere characterized by coherent scattering and introduces the concept of radiative-mixing equilibrium, which arises from energy transport via vertical mixing. 
In \S \ref{sec rte} we review the radiative transfer equation, the concept of radiative equilibrium (RE), and the boundary conditions pertinent to solving the coherent scattering atmosphere. 
When accounting for the energy flux resulting from vertical mixing, the concept of radiative equilibrium is modified to what we refer to as radiative-mixing equilibrium (RME), as presented in \S \ref{sec rme}. Finally, we discuss the approximation employed in our calculations in \S \ref{sec sga}. 

\subsection{Radiative transfer equation with coherent scattering}\label{sec rte}

In a steady-state, horizontally homogeneous atmosphere, the radiative transfer equation is given by \citep{1960Chandrasekhar,2017Hengbook}
\begin{equation}
    \mu\frac{\partial I_{\rm \nu}}{\partial \tau_{\rm \nu}}=I_{\rm \nu}-S_{\rm \nu} \ .
\end{equation}
Here, $\mu \equiv \cos \theta$, where $\theta$ represents the angle with respect to the vertical direction. $I_{\nu}$ is the wavelength-dependent intensity at frequency ${\nu}$, and $\tau_{\nu}$ represents the optical depth. 
The source function $S_{\nu}$ is defined as the ratio of total emissivity to total opacity, encompassing both scattering and thermal emission. 
 The complexity inherent to $S_{\nu}$,  as articulated by \citet{2015Hubeny} and \citet{stamnes2017}, renders the analytical solution of the radiative transfer equation difficult.
Consequently, a numerical approach is indispensable to accurately account for the myriad physical processes influencing radiative transfer accurately.

In the context of coherent scattering within the atmosphere, the radiative transfer equation for any given frequency $\nu$ is expressed as \citep{1960Chandrasekhar,1989Goody,2017Hengbook}:
\begin{equation}\label{governing eq}
    \mu\frac{\partial I_{\rm \nu}}{\partial \tau_{\rm \nu}}=I_{\rm \nu}-\frac{\omega_{\rm \nu}}{4\pi}\int_{\rm 0}^{4\pi}\mathcal{P}_{\rm \nu}I_{\rm \nu}d\Omega^{\prime}-(1-\omega_{\rm \nu})B_{\rm \nu} \ .
\end{equation}
Here $\mathcal{P}_{\nu}$ is the scattering phase function, and integrated over all incident angles in spherical coordinates, we have 
\begin{equation}
   \int_{\rm 0}^{4\pi}\mathcal{P}_{\rm \nu}d\Omega=4\pi \ .
\end{equation}
The single-scattering albedo is the ratio of the scattering cross-section to the total cross-section, i.e., $\omega_{\rm \nu} = \sigma_{\rm \nu} / (\kappa_{\rm \nu} + \sigma_{\rm \nu})$. $B_{\rm \nu}$ is the Planck function that describes the spectral radiance at a temperature $T$.
In addition, the three moments of the radiation intensity is defined as:
\begin{equation}
    J_{\nu} \ , \ H_{\rm \nu} \ , \ K_{\rm \nu}=\frac{1}{2}\int^{1}_{\rm -1}(1 \ ,\ \mu \ , \ \mu^2)I_{\rm \nu}(\mu)d\mu \ .
\end{equation}
Following \citet{Pierrehumbert2010},\citet{Heng2014} and \citet{Heng2017}, we multiply Equation~\eqref{governing eq} by a specific function $\mathcal{H}(\theta)$ and integrate, the result is shown as: 
\begin{align}
\label{governing eq1}
    \frac{\partial}{\partial\tau_{\rm \nu}}\int_{\rm 0}^{2\pi}\int_{\rm -1}^{1}\mu\mathcal{H}I_{\rm \nu}d\mu d\phi&=\nonumber\\
    \int_{\rm 0}^{2\pi}\int_{\rm -1}^{1}\mathcal{H}I_{\rm \nu}d\mu d\phi-\mathcal{I}-&(1-\omega_{\rm \nu}) \int_{\rm 0}^{2\pi}\int_{\rm -1}^{1}\mathcal{H}B_{\rm \nu}d\mu d\phi \ , 
\end{align}
where
\begin{align}
\label{governing eq2}
    \mathcal{I} =\omega_{\rm \nu} \int_{\rm 0}^{2\pi} \int_{\rm -1}^{1} \mathcal{G}I_{\rm \nu}d\mu d\phi \ ,\nonumber\\
    \mathcal{G}=\frac{1}{4\pi}\int_{\rm 0}^{2\pi}\int_{\rm -1}^{1}\mathcal{H}\mathcal{P}_{\rm \nu}d\mu d\phi \ .  
\end{align}
Given $\mathcal{H} = 1$, Equations~\eqref{governing eq2} follow that $\mathcal{G} = 1$ and $\mathcal{I} = 4\pi \omega_{\nu} J_{\nu}$. From Equation~\eqref{governing eq1}, the derivative of the first moment of radiation intensity in the case of non-isotropic scattering is:
\begin{equation}\label{governing1}
    \frac{\partial H_{\rm \nu}}{\partial\tau_{\rm \nu}}=(1-\omega_{\rm \nu})(J_{\rm \nu}-B_{\rm \nu}) \ .
    \end{equation}
For $\mathcal{H} = \mu$, we obtain (See \citealt{Heng2014} for details.)
\begin{equation}
\mathcal{I} =\omega_{\rm \nu}\mathrm{g}_{\rm 0\nu}H_{\rm \nu} \ .
\end{equation}
$\mathrm{g}_{\rm 0\nu}$ is the asymmetry factor which is defined as:
\begin{equation}\label{g0v}
    \mathrm{g}_{\rm 0\nu}\equiv\frac{1}{4\pi}\int_{\rm 0}^{4\pi}\mu\mathcal{P}_{\rm \nu}d\Omega \ .
\end{equation}

Together with Equation \eqref{governing eq1}, the second moment of the radiation intensity in the context of non-isotropic scattering is given by 
\begin{equation}\label{governing2}
    \frac{\partial K_{\rm \nu}}{\partial\tau_{\rm \nu}}=\gamma_{\rm \nu}H_{\rm \nu} \ ,
\end{equation}
where $\gamma_{\rm \nu}=(1-\omega_{\rm \nu}\mathrm{g}_{\rm 0\nu})$. By merging Equations \eqref{governing1} and \eqref{governing2}, a second-order equation of the intensity moments is derived:
\begin{equation}\label{eq.diffrential}
    \frac{\partial^2K_{\rm \nu}}{\partial\tau_{\rm \nu}^2}=\gamma_{\rm \nu}(1-\omega_{\rm \nu})(J_{\rm \nu}-B_{\rm \nu}) \ .
\end{equation}
Introducing the Eddington factor to substitute $K_{\nu}$ with $J_{\nu}$ is beneficial, i.e.,
\begin{equation}\label{Eddingdon 1}
    f_{\rm \nu}=\frac{K_{\rm \nu}}{J_{\rm \nu}}=\frac{\int^1_{\rm -1}I_{\rm\nu}(\mu)\mu^2 d\mu}{\int^1_{\rm -1}I_{\rm\nu}(\mu) d\mu} \ ,
\end{equation}
and Equation~\eqref{eq.diffrential} becomes
\begin{equation}\label{eq.RTE}
    \frac{\partial^2(f_{\rm \nu}J_{\rm \nu})}{\partial\tau_{\rm \nu}^2}=\gamma_{\rm \nu}(1-\omega_{\rm \nu})(J_{\rm \nu}-B_{\rm \nu}) \ .
\end{equation}
In contrast to Equation 3 in Paper I \citep{2024Zhong}, scattering introduces the coefficient term $\gamma_{\rm \nu}(1-\omega_{\rm \nu})$, and $\gamma_{\rm \nu}(1-\omega_{\rm \nu})=1$ when scattering is ignored. 
From Equation \eqref{governing2}, 
the upper boundary at the atmosphere's top is:
\begin{equation}\label{bc1}
    \left[\frac{\partial(f_{\rm \nu}J_{\rm \nu})}{\partial\tau_{\rm \nu}}\right]_{\rm \rm 
 \tau=0}=\gamma_{\rm \nu}\left[\mathrm{g}_{\rm \nu}J_{\rm \nu}(0)-H_{\rm \nu}^{\rm ext}\right] \ .
\end{equation}
$H_{\rm \nu}^{\rm ext}$ is the external irradiation at the top of the atmosphere. $\mathrm{g}_{\rm \nu}J_{\rm \nu}(0)$ is the outgoing flux, where $\mathrm{g}_{\rm \nu}$ is the surface Eddington factor and is shown as:
\begin{equation}\label{Eddingdon 2}
    \mathrm{g}_{\rm \nu}=\frac{H^+_{\rm \nu}(0)}{J_{\rm \nu}(0)}=\frac{\int^1_{\rm 0}I_{\rm\nu}(\mu,\tau=0)\mu d\mu}{\int^1_{\rm -1}I_{\rm\nu}(\mu,\tau=0) d\mu} \ .
\end{equation}
The numerator is computed over the interval from $0$ to $1$ to account for the outward flux. The numerical methodologies for solving $\mathrm{g}_{\nu}$ and $f_{\nu}$ are elaborated in \citet{Hubeny2017} and \citet{2000Sudarsky}. To comply with the diffusion approximation, the bottom boundary condition of the atmosphere is expressed by
\begin{equation}\label{bc2}
    \left[\frac{\partial(f_{\rm \nu}J_{\rm \nu})}{\partial\tau_{\rm \nu}}\right]_{\rm \tau_{\rm max}}=\gamma_{\rm \nu}\left[\frac{1}{2}(B_{\rm \nu}-J_{\rm \nu})+\frac{1}{3}\frac{\partial B_{\rm \nu}}{\partial\tau_{\rm \nu}}\right]_{\rm \tau_{\rm max}} \ .
\end{equation}
 The temperature profile is determined by the energy balance in each atmospheric layer, described by radiative equilibrium. The condition for RE is expressed as: 
 \begin{equation}\label{balance1}
     \int^\infty_{\rm 0}\kappa_{\rm \nu}(J_{\rm \nu}-B_{\rm \nu})d\nu=0 \ ,
 \end{equation}
and rewritten in differential form as:
  \begin{equation}\label{balance2}
     \int^\infty_{\rm 0}\frac{1}{\gamma_{\rm \nu}}\frac{d(f_{\rm \nu}J_{\rm \nu})}{d\tau_{\rm \nu}}d\nu=\frac{\sigma_{\rm R}}{4\pi}T^4_{\rm int} \ .
 \end{equation}
Here, 
$T_{\text{int}}$, the internal temperature, is commonly incorporated in 1D models to supply energy to the lower layers of the atmosphere. It is related to the total energy flux emanating from the planetary interior, which represents the residual heat from planetary formation. \cite{2011Fortney} estimated that $T_{\rm int}$ of Jupiter is about 99 K to match the observed effective temperature.  More recently, \cite{2024Welbanks} reported that WASP-107b has an internal temperature exceeding 345 K. Paper I demonstrates the heating effect of eddy flux on the deep atmosphere, indicating that when eddy flux is included in the calculations, a different value for $T_{\text{int}}$ may be obtained. As the planet cools, this internal heat progressively decreases. When atmospheric kinetic processes are intense, the contribution of internal heating becomes less important. 

\subsection{Radiative-mixing equilibrium}\label{sec rme}
Atmospheric circulation and gravity wave breaking contribute to vertical mixing in convectively stable layers. \citet{Youdin2010} examines the influence of additional heat flux resulting from this mixing process. Such mixing leads to entropy mixing, which facilitates heat transfer from lower to higher temperature regions, consequently increasing the temperature of the lower atmosphere \citep{Youdin2010, 2018Leconte}. The mixing flux is given by
\begin{equation}\label{eq.F}
    F_{\rm eddy}=-K_{\rm zz}\rho g\left(1-\frac{\nabla}{\nabla_{\rm ad}}\right) \ ,
\end{equation}
where the logarithmic temperature gradient is
\begin{equation}\label{nabla}
    \nabla= \frac{d\ln T }{d\ln P}=\frac{P}{T}\frac{dT}{dP} \ .
\end{equation}
 The adiabatic gradient for an ideal diatomic gas is $\nabla_{\rm ad} = 2/7$. The intensity of the mixing flux is determined by the eddy diffusion $K_{\rm zz}$. Its value is influenced by the atmospheric properties.  \citet{Youdin2010} identified a possible upper limit of $10^3$ to $10^5 \ \text{cm}^2 \ \text{s}^{-1}$ for stable hot Jupiters. While \citet{2021Blain} reported $K_{\rm zz}$ values for the sub-Neptune K2-18b, ranging from $10^6$ to $10^9 \ \text{cm}^2 \ \text{s}^{-1}$, and extended this range to $10^5$ to $10^{10} \ \text{cm}^2 \ \text{s}^{-1}$ to address uncertainties. 
The source of this energy flux is mainly from small-scale turbulent processes, which occur on scales much smaller than the pressure scale height. Such atmospheric motion resembles diffusion so we can use the eddy diffusion approximation and derive the formula of the flux that links to the local temperature gradient and density.  

Vertical mixing influences the RE through the induced energy flux, making it necessary to evolve RE to RME. The RME is defined as:
 \begin{equation}\label{nbalance1}
     \int^\infty_{\rm 0}\kappa_{\rm \nu}(J_{\rm \nu}-B_{\rm \nu})d\nu+\frac{ g}{4 \pi}\frac{d F_{\rm eddy}}{dP}=0 \ ,
 \end{equation}
  \begin{equation}\label{nbalance2}
     \int^\infty_{\rm 0}\frac{1}{\gamma_{\rm \nu}}\frac{d(f_{\rm \nu}J_{\rm \nu})}{d\tau_{\rm \nu}}d\nu+\frac{F_{\rm eddy}}{4\pi}=\frac{\sigma_{\rm R}}{4\pi}T^4_{\rm int} \ .
 \end{equation}
Following \citet{Gandhi2017}, we adopt Equation \eqref{nbalance2} for the lower atmosphere and Equation \eqref{nbalance1} for the upper atmosphere to enhance the numerical stability of our models.

\subsection{Semi-grey approximation}\label{sec sga}

This study investigates the influence of vertical mixing and scattering on the temperature-pressure profile of exoplanetary atmospheres. Although chemical composition plays a role in determining opacity, it is not the primary emphasis of this investigation.
We adopt the semi-grey approximation \citep{2010Guillot, Heng2014}, dividing opacity into two main components: the visible band (``v'') for incoming irradiation and the thermal band (``th'') for outgoing emission, with physical quantities integrated across these bands and labeled accordingly \citep{2010Guillot}. In this approach, frequency integrals in the radiative transfer equations are replaced by band-specific terms. We assume $B_{\rm v}=0$ and $B_{\rm th}=B=\sigma_{\rm R}T^4/\pi$, with $\kappa_{\rm \nu}$, $\omega_{\rm \nu}$, and $\mathrm{g}_{\rm 0\nu}$ treated as constants.

To ensure consistency with previous studies, we adopt the Eddington factor proposed by \citet{Heng2014}. This approach is outlined as follows
\begin{equation}\label{eq fth}
    f_{\rm th}=\frac{K_{\rm th}}{J_{\rm th}}=\frac{1}{3} \ , \ \mathrm{g}_{\rm th}=\frac{H_{\rm th}(0)}{J_{\rm th}(0)}=\frac{3}{8} \ ,
\end{equation}
\begin{equation}\label{eq fv}
    f_{\rm v}=\frac{K_{\rm v}}{J_{\rm v}}=\mu_{\rm *}^2 \ , \ \mathrm{g}_{\rm v}=\frac{H^+_{\rm v}(0)}{J_{\rm v}(0)}=0 \ .
\end{equation}
Here, $H^+_{\rm th}(0)=H_{\rm th}(0)$ because the incoming irradiation is assumed to be confined solely to visible bands.
The cosine of the irradiation angle, $\theta_{\rm *}$, is defined as $\mu_{\rm *} \equiv \cos \theta_{\rm *}$.
In addition, $\mathrm{g}_{\rm v}=0$ denotes the absence of visible flux in the outgoing radiation at the atmosphere's surface. As the temperatures within planetary atmospheres are considerably lower than the effective temperature of the host star, the radiation fields are largely decoupled, which supports the appropriateness of the semi-grey approximation. The discrepancy between the value of Eddington factor $\mathrm{g}_{\rm v}$ in this paper and that reported in our Paper one arises because only the outward direction of the visible band is considered here. 
Additionally, the value of $\mathrm{g}_{\rm th}$ also differs from that in Paper one
as $\mathrm{g}_{\rm th}=1/2$ is inconsistent with the other Eddington coefficients and should be changed to $3/8$ as described in \cite{Heng2014}and \cite{2017Hengbook}.

To formulate the temperature-pressure profile, we transform the independent variable from opacity to atmospheric pressure. In hydrostatic equilibrium, we have
\begin{equation}\label{eq t-p}
    d\tau_{\rm \nu}=\frac{\kappa_{\rm \nu} + \sigma_{\rm \nu}}{g}dP \ ,
\end{equation}
We assume a constant gravitational acceleration, $g=10^3cm/s^{2}$ and $P = mg$. The absorption and scattering coefficients are expressed as opacities per unit mass, measured in $cm^2/g$.

We assume a constant value for $K_{\rm zz}$ without performing an extensive variation analysis to investigate the combined impacts of mixing flux and scattering on the $T$-$P$ profile. 
Since the Planck function $B_{\rm \nu}$ and mixing flux $F_{\rm eddy}$ are the nonlinear functions of temperature $T$, deriving a definitive analytic result presents significant challenges. 
To obtain the temperature profile by solving the nonlinear Equations (\ref{eq.RTE}, \ref{nbalance1}, \ref{nbalance2}), we employ numerical methods and linearization techniques. These approaches are elaborated in the literature, including works by \cite{Gandhi2017} and \cite{Hubeny2017}.
Some details regarding our calculations are presented in Appendix  \ref{Linearization}.
\subsection{Temperature profile with both scattering and Mixing flux}\label{sec Analytic solution} 

Before presenting the numerical results, we introduce an analytic solution for an atmosphere with vertical mixing and non-isotropic scattering in this section.
Following \cite{Heng2014}, we derive the relationship between atmospheric temperature and mixing flux ($F_{\rm eddy}$) using the semi-grey approximation. 
We defined the parameters $\beta_{\nu}$ at different frequency band $\nu$ as:
\begin{equation}
  \beta_{\rm \nu}=\frac{P}{g}\frac{\kappa_{\rm \nu}}{\beta_{0\rm \nu}} \ ,
\end{equation}
with
\begin{equation}\label{eq.beta0}
    \beta_{\rm 0\nu}=\left(\frac{1-\omega_{\rm \nu }}{\gamma_{\rm \nu}}\right)^{1/2} \ .
\end{equation}
In combination with Equation \eqref{eq t-p}, Equation \eqref{eq.RTE} becomes:
\begin{align}
     \frac{\partial^2(f_{\rm \nu}J_{\rm \nu})}{\partial P^2}&=\frac{\gamma_{\rm \nu}}{g^2}(\kappa_{\rm \nu} + \sigma_{\rm \nu})^2(1-\omega_{\rm \nu})(J_{\rm \nu}-B_{\rm \nu}) \nonumber\\
     &=\frac{\gamma_{\rm \nu}}{g^2}\frac{\kappa_{\rm \nu}^2}{1-\omega_{\rm \nu}}(J_{\rm \nu}-B_{\rm \nu})
     \ .
\end{align}
In the visible band, we have $B_{\rm v}=0$ and $f_{\rm v}=\mu_{\rm *}^2$, so the equation becomes:
\begin{equation}\label{eq s}
    \frac{\partial^2J_{\rm v}}{\partial P^2}=\left(\frac{\kappa_{\rm v}}{g\beta_{\rm 0v}\mu_{*}}\right)^2J_{\rm v} \ .
\end{equation} 
Its solution satisfies 
\begin{equation}\label{eq Jv}
    J_{\rm v}=J_{\rm v}(0)\exp\left(\frac{\beta_{\rm v}}{\mu_{*}}\right) \ .
\end{equation}
Here, the visible opacity $\kappa_{\rm v}$ is considered to be constant.
 In combination with Equation \eqref{eq t-p}, Equation \eqref{governing1} in visible band becomes:
\begin{equation}\label{eq.H/P}
    \frac{\partial H_{\rm v}}{\partial P}=\frac{\kappa_{\rm v}}{g}J_{\rm v} \ .
\end{equation}
With the expression for $J_{\rm v}$, we integrate Equation \eqref{eq.H/P} and then obtain the first moment in the visible band, $H_{\rm v}$, that is:
\begin{equation}\label{eq Hv}
    H_{\rm v}=H_{\rm v}(0)\exp\left(\frac{\beta_{\rm v}}{\mu_{*}}\right) \ .
\end{equation}
with $H_{\rm v}(0)=J_{\rm v}(0)\beta_{\rm 0v}\mu_*$. 
The radial vector of the planet, oriented upwards, is defined with a zero-degree angle. As a result, the parameter $\mu_{*}$, related to downward irradiation, is negative. 
To ensure that $J_{\rm v}$ and $H_{\rm v}$ approach zero as $P\rightarrow \infty$, the solution excludes the negative exponent branch.

 When transforming the independent variable from opacity to atmospheric pressure and using the Eddington factor, Equation~\eqref{governing2} becomes:
\begin{equation}\label{ngoverning2}
    \frac{\partial (f_{\rm \nu}J_{\rm \nu})}{\partial P}=\frac{\gamma_{\rm \nu}\kappa_{\rm \nu}}{
    (1-\omega_{\rm \nu})g}H_{\rm \nu} \ ,
\end{equation}

Integrating Equation~\eqref{ngoverning2} along pressure in the thermal band, we obtain:
\begin{equation}\label{eq. T1}
    J_{\rm th}- J_{\rm th}(0)=\frac{1}{f_{\rm th}}\int_0^P\frac{\kappa_{\rm th}}{g\beta_{\rm 0th}^2}H_{\rm th}dP^{\prime} \ .
\end{equation}
In RME, Equation~\eqref{nbalance2} implies that the first moment of the specific intensity obeys: 
\begin{equation}\label{eq.H1}
    H=H_{\rm th}+H_{\rm v}+\frac{F_{\rm eddy}}{4\pi}=\frac{\sigma_{\rm R}T_{\rm int}^4}{4\pi} \ .
\end{equation}
Thus, Equation~\eqref{eq. T1} becomes:
\begin{equation}\label{eq Jth}
    J_{\rm th}- J_{\rm th}(0)=\frac{1}{f_{\rm th}}\int_0^P\frac{\kappa_{\rm th}}{g\beta_{\rm 0th}^2}\left(\frac{\sigma_{\rm R}T_{\rm int}^4}{4\pi}-H_{\rm v}-\frac{F_{\rm eddy}}{4\pi}\right)dP^{\prime} \ .
\end{equation}
Under the semi-grey approximation, the energy conservation (Equation \ref{nbalance1}) has the form as:
\begin{equation}
    \kappa_{\rm th}J_{\rm th}-\kappa_{\rm th}B+\kappa_{\rm v}J_{\rm v}+\frac{ g}{4 \pi}\frac{d F_{\rm eddy}}{dP}=0 \
\end{equation}
With Equation \eqref{eq Jv}, the Planck function can be expressed as:
\begin{equation}\label{eq B}
 B=J_{\rm th}+\frac{\kappa_{\rm v}}{\kappa_{\rm th}}J_{\rm v}(0)\exp\left(\frac{\beta_{\rm v}}{\mu_{*}}\right)+\frac{ g}{4 \pi \kappa_{\rm th}}\frac{d F_{\rm eddy}}{dP}
\end{equation}
At the top of the atmosphere, $H_{\rm th}(0)=H-H_{\rm v}(0)-F_{\rm eddy}(0)/(4\pi)$, $J_{\rm th}(0)=H_{\rm th}(0)/\mathrm{g}_{\rm th}$ and the incoming flux satisfies $H_{\rm v}(0)=\mu_{*}\sigma_{\rm R}T_{irr}^4/(4\pi)$. 
By combining with the Equation \eqref{eq Hv} and the Equation \eqref{eq Jth}, the formulation presented in Equation \eqref{eq B} finally changes into:
\begin{align}\label{eq. nanalytic}
    &T^4=\frac{T_{\rm int}^4}{4}\left(\frac{1}{\mathrm{g}_{\rm th}}+\frac{\kappa_{\rm th}P}{f_{\rm th}\beta_{\rm 0th}^2g}\right)\nonumber\\
    &+\frac{T_{\rm irr}^4}{4}\left[-\frac{\mu_{\rm *}}{\mathrm{g}_{\rm th}}+\frac{\kappa_{\rm v}}{\kappa_{\rm th}}\frac{1}{\beta_{\rm 0v}}e^{\frac{\beta_{\rm v}}{\mu_{\rm *}}}-\frac{\kappa_{\rm th}}{\kappa_{\rm v}}\frac{\beta_{\rm 0v}\mu_{\rm *}^2}{f_{\rm th}\beta_{\rm 0th}^2}(e^{\frac{\beta_{\rm v}}{\mu_{\rm *}}}-1)\right]\nonumber\\
    &+\frac{1}{4\sigma_{\rm R}}\left[\frac{g}{\kappa_{\rm th}}\frac{d F_{\rm eddy}}{d P}-\frac{F_{\rm eddy}(0)}{\mathrm{g}_{\rm th}}-\frac{\kappa_{\rm th}}{gf_{\rm th}\beta_{\rm 0th}^2}\int_0^PF_{\rm eddy}dP^{\prime} \right] \ . 
\end{align}
Changing $F_{\rm eddy}$ to $4\pi \int^{\infty}_{\rm m} q\nabla T dm$, we obtain the same expression as Equation (43) of \cite{2010Guillot} when ignoring scattering. 
Replacing $-\left(g/4\pi\right)\partial F_{\rm eddy}/\partial P$ with the heating rate $Q$ makes Equation \eqref{eq. nanalytic} consistent with \cite{Heng2014}, when assuming that heating at the bottom is included in the term related to $T_{\rm int}$. 
In addition, Equation \eqref{eq. nanalytic} can be simply split into two parts: 
\begin{equation}
    T^4=j_{\rm r}+j_{\rm F} \ .
\end{equation}
$j_{\rm F}$ is the third part of the Equation \eqref{eq. nanalytic} and represent the direct effect of $F_{\rm eddy}$.
$F_{\rm eddy}(0)$ 
vanishes because the density ($\rho$) equals zero at the top of the atmosphere. Therefore, only the integration and differentiation of $F_{\rm eddy}$ plays a role in the temperature. In the upper atmosphere, the low density makes $F_{\rm eddy}$ too small to have much effect on temperature. 
Moreover, $j_{\rm F}$ is expressed as:
\begin{equation}\label{jF}
    j_{\rm F}=\frac{1}{4\sigma_{\rm R}}\left[\frac{g}{\kappa_{\rm th}}\frac{\partial F_{\rm eddy}}{\partial P}-\frac{\kappa_{\rm th}}{gf_{\rm th}\beta_{\rm 0th}^2}\int_0^PF_{\rm eddy}dP^{\prime} \right] \ .
\end{equation}
As pressure increases, the integral term becomes dominant over the differential term. In regions of high pressure, the effect of the eddy flux on temperature is primarily determined by the magnitude of the flux integral, which demonstrates the cumulative effect of the energy transport induced by mixing.
\par

\section{Result}\label{result}
We explore the interaction between mixing flux and scattering using two approaches.
First, we vary $K_{\rm zz}$ from $0$ to $10^9 \ \text{cm}^2 \ \text{s}^{-1}$ to observe its impact on both scattering greenhouse and anti-greenhouse effects. Then, with $K_{\rm zz}$ fixed at $10^6 \ \text{cm}^2 \ \text{s}^{-1}$, we adjust $\sigma_{\rm \nu}$ to see how scattering affects vertical mixing. Our findings highlight a mutual influence between vertical mixing and scattering on the strength of greenhouse and anti-greenhouse effects. 
The greenhouse effect has complicated, non-linear effects on the atmospheres of planets. In this work, we did consider only the heating impact of the greenhouse effect, and the localized cooling effect of the anti-greenhouse effect. A reduction in atmospheric warming is referred to as a decrease in the greenhouse effect, while a reduction in cooling is termed a decrease in the anti-greenhouse effect.
The numerical results are detailed in $\S$ \ref{sec long}, with temperature inversion outcomes in $\S$ \ref{sec inversion}.


\subsection{Interplay between scattering and mixing flux}\label{sec long}
Scattering can either increase or decrease atmospheric temperatures, thereby contributing to the greenhouse or anti-greenhouse effects \citep{Heng2014}. 
Vertical mixing produces a \textit{mechanical greenhouse effect} \citep{Youdin2010}, which generally results in elevated temperatures.
The interaction between scattering and vertical mixing is inherently non-linear. Specifically, scattering alters the thermal structure of the atmosphere, leading to fluctuations in the mixing flux $F_{\rm eddy}$ and variable temperature responses.
Furthermore, Equation \eqref{jF} illustrates a direct coupling between thermal scattering and $F_{\rm eddy}$.
Variations in $\beta_{\rm 0th}$ directly impact the influence of $F_{\rm eddy}$ on temperature. \par
\begin{figure}
\includegraphics[width=\columnwidth]{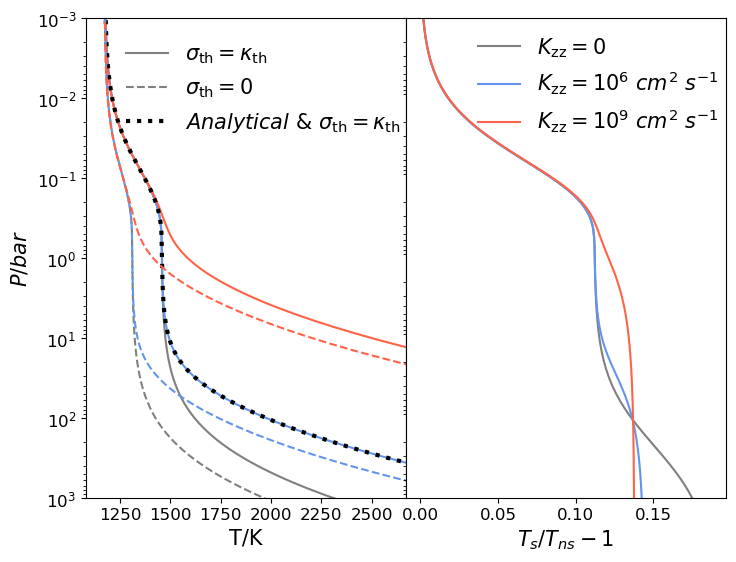}
\centering
    \caption{The left panel displays the temperature profiles with and without thermal isotropic scattering, varying across different values of $K_{\rm zz}$. Solid lines denote the cases where $\sigma_{\rm th}=\kappa_{\rm th}=0.01 \ \text{cm}^2\ \text{g}^{-1}$ , while dashed lines represent the case with $\sigma_{\rm th}=0$. Each color corresponds to a distinct value of $K_{\rm zz}$. Notably, the dashed gray line illustrates the condition devoid of both vertical mixing and scattering. The black dotted line depicts the quasi-analytical solution derived from Equation \eqref{eq. nanalytic} with parameters $\sigma_{\rm th}=\kappa_{\rm th}$ and $K_{\rm zz}=10^6  \ \text{cm}^2\ \text{s}^{-1}$. The values of $F_{\rm eddy}$ in Equation \eqref{eq. nanalytic} are sourced from numerical calculations. The right panel illustrates the corresponding fractional change in temperature due to scattering, 
    $T_{\rm s}/T_{\rm ns}-1$, across varying values of $K_{\rm zz}$. The parameters are set as $g = 10^3 \ \text{cm} \ \text{s}^{-2}$, $T_{\rm int} = 200 \ \text{K}$, and $T_{\rm irr} = 1200 \ \text{K}$.}
    \label{fig xl+Fchange}
\end{figure}
The greenhouse effect resulting from thermal scattering is influenced by variations in vertical mixing, as illustrated in Figure \ref{fig xl+Fchange}. 
This influence is quantified by the ratio of the scattering temperature to the non-scattering temperature, $T_{\rm s}/T_{\rm ns}$.
In the left panel, solid lines represent temperature profiles with scattering, while dashed lines depict profiles without scattering across different mixing flux strengths. 
The numerical solutions align well with the analytical results presented in Section \ref{sec Analytic solution}, thereby confirming the validity of our simulation. 
Specifically, this alignment is evident in the black dotted lines corresponding to $K_{\rm zz}=10^6  \ \text{cm}^2\ \text{s}^{-1}$. 
The right panel displays the values of $T_{\rm s}/T_{\rm ns}-1$. 
As the mixing strength increases, this ratio diminishes in the lower atmosphere, indicating a reduction in the scattering-induced greenhouse effect. 
Conversely, in the middle layers, $T_{\rm s}/T_{\rm ns}-1$ increase with $K_{\rm zz}$.\par


\begin{figure}
\includegraphics[width=\columnwidth]{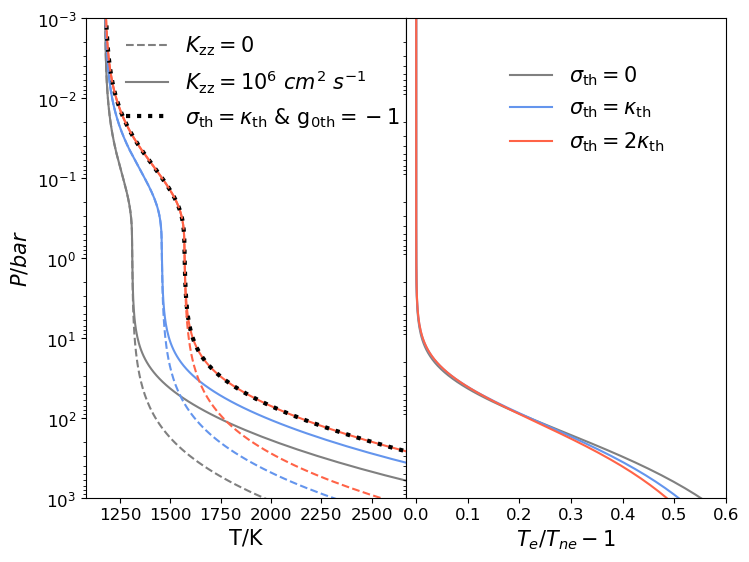}
    \caption{The left panel presents the temperature profiles both with and without vertical mixing, for various values of $\sigma_{\rm th}$. The solid lines represent a vertical mixing coefficient of $K_{\rm zz} = 10^6\ \text{cm}^2\ \text{s}^{-1}$, while dashed lines correspond to the scenario where $K_{\rm zz} = 0$. 
    The black dotted line reflects the quasi-analytical solution obtained from Equation \eqref{eq. nanalytic} with parameters set to $\sigma_{\rm th} = \kappa_{\rm th}$, $K_{\rm zz} = 10^6\ \text{cm}^2\ \text{s}^{-1}$ and the asymmetry factor $\mathrm{g}_{\rm 0th} = -1$. 
    The right panel illustrates the fractional change in temperature 
    due to vertical mixing, $T_{\rm e}/T_{\rm ne}-1$, for each value of $\sigma_{\rm th}$. 
    The other parameters remain consistent with those presented in Figure \ref{fig xl+Fchange}.}
    \label{fig F+xlchange}
\end{figure}
Thermal scattering modifies the greenhouse effect from vertical mixing. 
This modification is quantified by the change in the fractional change in temperature due to mixing flux, 
expressed as $T_{\rm e}/T_{\rm ne}-1$. 
Figure \ref{fig F+xlchange} illustrates these findings across different scattering coefficient values.
In the middle atmosphere, $T_{\rm e}/T_{\rm ne}-1$ slightly increases with the increasing vertical mixing, although this is not easy to distinguish in the Figure.
The greenhouse effect attributed to mixing flux diminishes with increasing thermal scattering in the lower atmosphere. 
 An additive effect exists between thermal scattering and eddy flux, with their heating effects demonstrating a non-linear interaction.  Specifically, the combined effect in the lower atmosphere is less than the sum of their individual effects, as described by the inequality $1+1<2$.
As shown in Equation \eqref{eq.H1}, the eddy flux alters the average thermal radiation flux $H_{\rm th}$, which determines the influence of thermal scattering on temperature. In the absence of flux in the thermal band, thermal scattering has no influence on temperature. On the other hand, thermal scattering alters the temperature, which in turn influences the intensity of the eddy flux. As a result, the two phenomena mutually influence each other, leading to a non-linear collective effect.

It is natural to go ahead to the more general case, the non-isotropic scattering.  
Under the assumptions established in this study, each individual non-isotopic scattering value has an equivalent isotropic value when calculating the temperature profile. The equivalent isotropic value can be calculated as follows. 
Equation \eqref{eq. nanalytic} reveals that the temperature is a function of the parameter $\beta_{\rm 0\nu}$, rather than being directly influenced by the coefficients $\mathrm{g}_{\rm 0\nu}$ and $\omega_{\rm \nu}$ (or $\sigma_{\rm \nu}$). 
The degeneracy of $\beta_{\rm 0\nu}$, as presented in Equation \eqref{eq.beta0} suggests that a modification in $\mathrm{g}_{\rm 0\nu}$ may yield results equivalent to a specific value of $\omega_{\rm \nu}$ when $\mathrm{g}_{\rm 0\nu}=0$. 
The relationship between the single-scattering albedo of non-isotropic scattering and its isotropic counterpart can be described by the following equation:
\begin{equation}\label{eq. i-n}
    \omega_{\rm \nu}=\omega^{\prime}_{\rm \nu}\frac{1-\mathrm{g}^{\prime}_{\rm 0\nu}}{1-\omega^{\prime}_{\rm\nu}\mathrm{g}^{\prime}_{\rm 0\nu}} \ ,
\end{equation}
where $\omega^{\prime}_{\rm\nu}$ represents the single-scattering albedo for the non-isotropic scattering characterized by $\mathrm{g}^{\prime}_{\rm 0\nu}$.\par 
Equation \eqref{eq. i-n} demonstrates that a more backward scattering results in a temperature profile comparable to that produced by stronger isotropic scattering. 
The black dotted line in Figure \ref{fig F+xlchange} illustrates the temperature profile of a non-isotropic scattering atmosphere with {$\sigma^{\prime}_{\rm th}=\kappa_{\rm th}$} and $\mathrm{g}^{\prime}_{\rm 0th}=-1$. 
This profile aligns with the result of isotropic scattering with $\sigma_{\rm th}=2\kappa_{\rm th}$, as indicated in Equation \eqref{eq. i-n}, thus corroborating our analysis.
Moreover, in the scenario of pure forward scattering {\bf ($\mathrm{g}^{\prime}_{\rm 0\nu}=1$)}, the results obtained are equivalent to those in the absence of scattering ($\omega_{\rm \nu}=0$), since $\beta_{\rm \nu 0}=1$ applies to both scenarios.
Under conditions of pure forward scattering, the radiated light maintains its propagation direction and remains unaltered, effectively equating to the absence of scattering.
Given this equivalence, the subsequent analysis will concentrate exclusively on the results derived from isotropic scattering.\par
 Different temperature structures in the atmosphere, such as the temperature difference from the day side to the night side due to the different incident light angles of the star, will affect the results. The temperature difference will bring about the difference in chemistry and clouds which changes the strength of scattering. Quantifying the change in scattering is a cumbersome process, which is beyond the scope of our study. 

\begin{figure}
\includegraphics[width=\columnwidth]{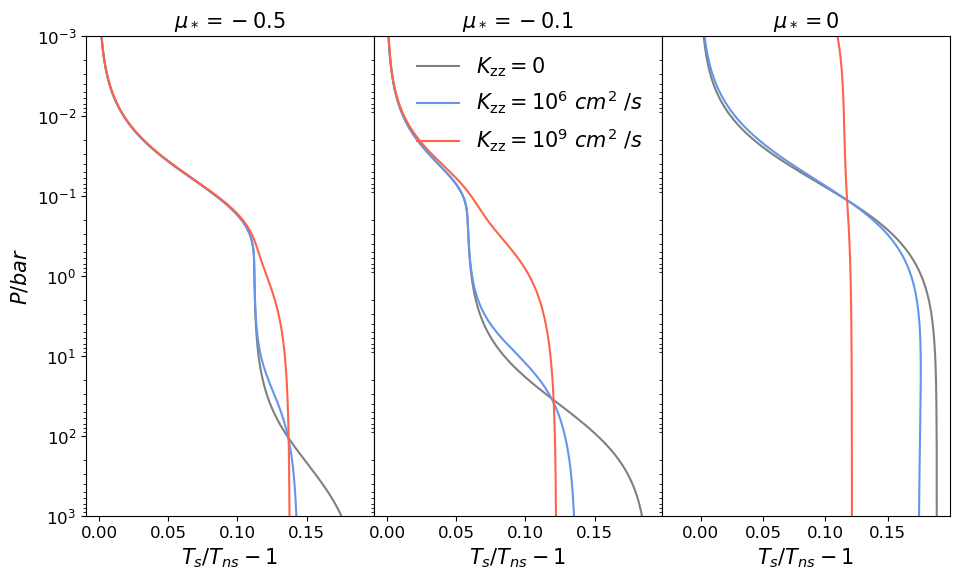}
    \caption{ These panels illustrate the fractional change in temperature due to thermal scattering, $T_{\rm s}/T_{\rm ns}-1$, for each value of $K_{\rm zz}$. $\mu_*$ is the cosine of the irradiation angle. $\mu_*=0$ is the night side.
    The other parameters are consistent with Figure 1 in the paper.}
    \label{fig Fmu+xlchange}
\end{figure}

We explored the results under different temperature structures from the day side to the night side by changing the parameter $\mu_*$ and continued to use parameterized methods to deal with scattering.
The results are shown in Figure \ref{fig Fmu+xlchange}. 
We get results similar to those for the substellar point. The difference lies in the positions of the switch between an increasing ratio and a diminishing ratio.
Near the night side, the atmospheric temperature weakens and the flux of radiation in the atmosphere becomes smaller, which makes the eddy flux relatively more significant. The integration of $F_{\rm eddy}$ at a lower pressure is sufficient to produce a noticeable effect so that the switch point occurs at a lower pressure.
\begin{figure}
\includegraphics[width=\columnwidth]{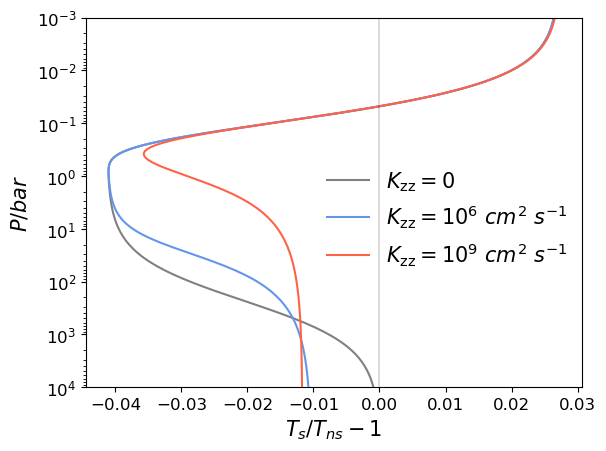}
    \caption{The fractional change in temperature
    induced by visible scattering across varying values of $K_{\rm zz}$. The parameter $\sigma_{\rm th}$ is set to zero, and  $\sigma_{\rm v}=\kappa_{\rm v}=\kappa_{\rm th}=0.01 \ \text{cm}^2\ \text{g}^{-1}$. All other parameters align with those indicated in Figure \ref{fig xl+Fchange}.}
    \label{fig xs}
\end{figure} 

The interplay between visible scattering and mixing flux is also examined, revealing minor adjustments.
Visible scattering generally cools the atmosphere, thus contributing to an anti-greenhouse effect \citep{Heng2014}. 
In the presence of mixing flux, the cooling in the middle atmospheric layers is diminished, while it is enhanced in the lower layers. These changes become more pronounced with increasing values of $K_{\rm zz}$,  as illustrated in Figure \ref{fig xs}.
 Although visible scattering influences the greenhouse effect of mixing flux, the impact is minimal and is not the focus of this paper.
\begin{figure}
\centering
\includegraphics[width=\columnwidth]{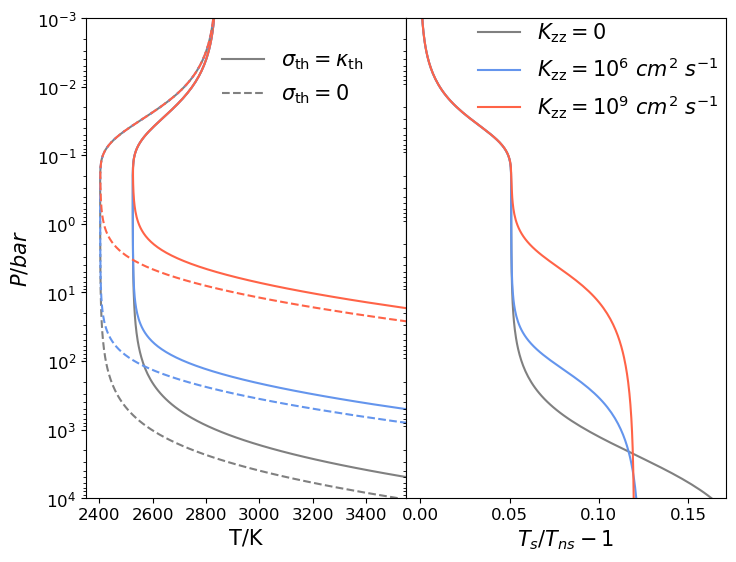}
    \caption{This figure is analogous to Figure \ref{fig xl+Fchange}, but illustrating an atmosphere with a temperature inversion. The left panel displays the temperature profiles with and without thermal isotropic scattering, varying across different values of $K_{\rm zz}$. The right panel illustrates the corresponding fractional change in temperature 
    due to scattering. The parameters are as follows: $\sigma_{\rm v}=0$, $\kappa_{\rm v} =0.04 \ cm^2 \ g^{-1}$, $\sigma_{\rm th}=\kappa_{\rm th}= 0.01 \ cm^2 \ g^{-1}$ and $T_{\rm irr}=2500K$.}
    \label{fig ks=4kl, xl}
\end{figure}

\subsection{Atmospheres with temperature Inversion }\label{sec inversion}
This section investigates the interaction between scattering and vertical mixing in the atmospheres with temperature inversions. We observed similar nonlinear patterns to those found in the atmospheres without temperature inversions.
Temperature inversions are induced by strong absorbers in the optical band.
Specifically, planets exposed to intense stellar irradiation can reach temperatures sufficient to sustain the gas-phase stability of metal oxides, such as $TiO$ and $VO$ \citep{2002Lodders}.  
Previous research has demonstrated that these metal oxides significantly modify the temperature structure of planetary atmospheres, creating pronounced thermal inversions at low pressures while cooling the deeper atmospheric layers \citep{2003Hubeny, 2015Parmentier}. 
To generate a temperature inversion, the visible opacity is raised to $0.04 \ cm^2 \ g^{-1}$, and the effective irradiation temperature is changed to be $T_{\rm irr}=2500K$. \par 
Figure \ref{fig ks=4kl, xl} illustrates how $K_{\rm zz}$ affects the greenhouse effect associated with thermal scattering. Figure \ref{fig /ks=4kl, xlF} illustrates the impact of thermal scattering on the greenhouse effect driven by mixing flux. The results obtained resemble those found in the atmosphere without temperature inversion. This similarity is attributed to the fact that temperature inversions primarily occur at low pressures, whereas the mixing flux predominantly influences temperature within the middle and lower atmospheric layers in our calculations. 
\par
\begin{figure}
\includegraphics[width=\columnwidth]{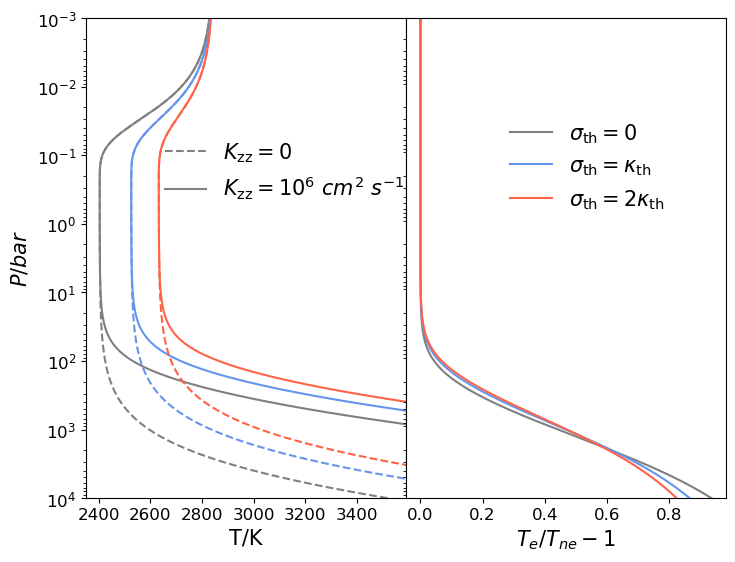}
    \caption{This figure is similar to Figure \ref{fig F+xlchange}, but for the atmosphere that has the temperature inversion. The left panel presents both the temperature profiles with and without vertical mixing, for various values of $\sigma_{\rm th}$. The right panel illustrates the fractional change in temperature 
    due to vertical mixing. The parameters are as follows: $\sigma_{\rm v}=0$, $\kappa_{\rm v} =0.04 \ cm^2 \ g^{-1}$, $\kappa_{\rm th}= 0.01 \ cm^2 \ g^{-1}$ and $T_{\rm irr}=2500K$. }
    \label{fig /ks=4kl, xlF}
\end{figure}
The vertical mixing flux decreases as temperature increases at constant pressure, attributable to its proportional relationship with atmospheric density.
Enhanced thermal scattering raises temperatures, concurrently decreasing the absolute magnitude of $F_{\rm eddy}$. 
As shown in Equation \eqref{jF} and Equation \eqref{eq.beta0}, stronger thermal scattering results in a smaller $\beta_{\rm 0th}$, which in turn amplifies the influence of $F_{\rm eddy}$ on temperature. 
The interplay of these two factors delineates how thermal scattering modifies the effect of $F_{\rm eddy}$ on the temperature.

If we only consider the mixed energy flux and radiation effects, in the upper atmosphere, the lower atmospheric density results in diminished flux,  thereby attenuating the effect of vertical mixing on temperature. However, in fact, the upper atmosphere is a dynamically active region, and the properties of this region are profoundly shaped by horizontal mixing of energy and matter transport, vertical mixing of matter transport processes, and photochemical processes, which are beyond the scope of this paper.
Within the middle layers, the temperature differences between scenarios with and without vertical mixing are relatively minor, yielding comparable values of $F_{\rm eddy}$.  Increased thermal scattering correlates with a smaller $\beta_{\rm 0th}$, which permits a consistent value of $F_{\rm eddy}$ to exert a more pronounced influence on temperature shifts (Equation \ref{jF}), thereby enhancing the fractional change in temperature 
in the middle atmosphere.
Conversely, in the lower atmospheric layers, significant variations in the integral of $F_{\rm eddy}$ across differing scattering conditions counterbalance and far outweigh the effects of a smaller $\beta_{\rm 0th}$ that arise from stronger thermal scattering. Consequently, the fractional change in temperature in the lower atmosphere decreases with the enhancement of thermal scattering.
These analyses also explain Figures \ref{fig F+xlchange} and \ref{fig /ks=4kl, xlF}. As $K_{\rm zz}$ increases, the influence of $j_{\rm F}$ on temperature becomes more significant, leading to greater modifications in the temperature ratio between scattering and non-scattering, as illustrated in Figure \ref{fig xl+Fchange} and Figure \ref{fig ks=4kl, xl}.

The third term in Equation \eqref{eq. nanalytic} does not explicitly incorporate visible scattering. Instead, it predominantly affects $j_{\rm F}$ by altering the value of $F_{\rm eddy}$. Visible scattering plays an important role in the low pressure region, but $F_{\rm eddy}$ is too small in this region to bring about significant changes in temperature. So our discussion is still focused on the middle and lower atmosphere. In the middle and lower atmosphere, the small temperature variations induced by visible scattering, as depicted in Figure \ref{fig ks=4kl, xlF}, lead to a negligible difference in the magnitude of $ \lvert F_{\rm eddy}\rvert$ between scenarios with and without scattering. 
As a result, these variations yield smaller modifications to the temperature ratio relative to those arising from thermal scattering.\par
\begin{figure}
\includegraphics[width=\columnwidth]{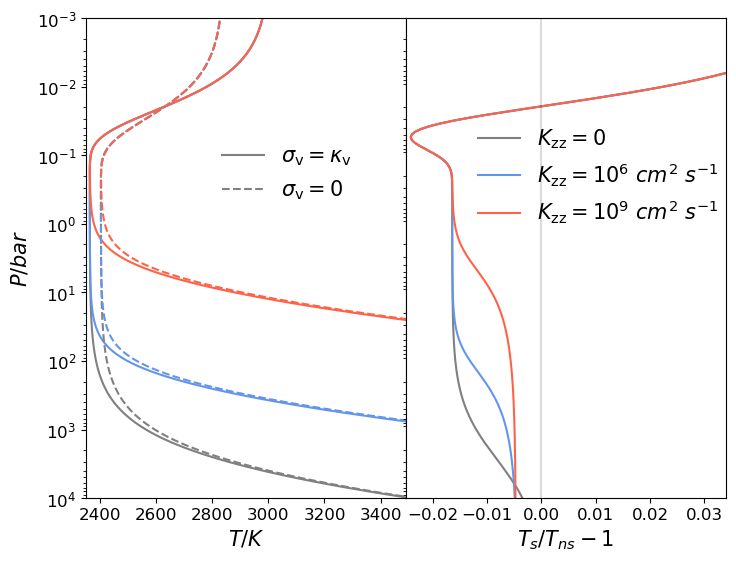}
    \caption{This figure is analogous to Figure \ref{fig ks=4kl, xl}, but showing the case of visible scattering. The left panel displays the temperature profiles with and without scattering, varying across different values of $K_{\rm zz}$. The right panel illustrates the corresponding fractional change in temperature
    due to scattering. The parameter $\sigma_{\rm th}$ is set to be zeros. All other parameters align with those indicated in Figure \ref{fig ks=4kl, xl}.}
    \label{fig ks=4kl, xlF}
\end{figure}
 Figure \ref{fig ks=4kl, xlF} depicts how mixing flux affects the anti-greenhouse effect associated with the visible scattering in atmospheres. In the middle atmosphere, visible scattering leads to a reduction in temperature, resulting in a larger absolute value of $F_{\rm eddy}$. Consequently, mixing flux induces a greater temperature increase in the case with visible scattering, thereby increasing $T_{\rm s}/T_{\rm ns}$.
In deeper atmospheric layers, the temperature difference between scattering and non-scattering scenarios gradually diminishes. Nonetheless, the temperature gradient in the non-scattering case is slightly greater than that in the scattering scenario, as implied in Figure \ref{fig ks=4kl, xlF}. Consequently, the absolute value of $F_{\rm eddy}$ in the non-scattering case becomes larger, and the integral of $F_{\rm eddy}$ gradually surpasses that in the scattering case, indicating that mixing flux induces a more pronounced temperature increase in atmospheres without scattering. Thus, in the lower layers, the vertical mixing case displays a smaller value of $T_{\rm s}/T_{\rm ns}$ than the non-mixing case. This also explains the results in Figure \ref{fig xs}.\par
\section{Conclusion}\label{conclusion}
This study employs numerical methods to investigate the interplay between scattering and mixing fluxes that arise from vertical mixing, and we compare these results with the analysis. To quantify the greenhouse and anti-greenhouse effects associated with scattering and vertical mixing, we calculate the ratio of scattering temperature to non-scattering temperature (i.e., $T_{\rm s}/T_{\rm ns}$) and the ratio of atmospheric temperature with mixing flux to that without mixing flux (i.e., $T_{\rm e}/T_{\rm ne}$). Our findings indicate that mixing flux and scattering alter the intensity of each other's greenhouse or anti-greenhouse effects on the atmosphere. Additionally, we provide a comparison of the solutions between isotropic and non-isotropic scattering scenarios under the semi-grey approximation and Eddington approximation.\par
The primary results of this study are as follows:
\begin{enumerate}
      \item Following \citet{Heng2014}, we derive the temperature formula of the coherent scattering atmosphere incorporating an additional mixing flux, as outlined in Equation \eqref{eq. nanalytic}. 
      \item Under the semi-grey and Eddington approximation, non-isotropic scattering can be effectively approximated by specific isotropic scattering when calculating the temperature profile, as demonstrated in Equation \eqref{eq. i-n}. A more backward scattering would work like a stronger isotropic scattering, bringing about a stronger interaction with the eddy flux, while a more forward scattering would do the opposite.
      \item For larger values of  $K_{\rm zz}$, the greenhouse effect resulting from thermal scattering is weaker in the lower atmosphere, whereas the anti-greenhouse effect associated with visible scattering is slightly stronger. In the middle atmosphere, an increase in the mixing flux enhances the warming effect from thermal scattering while concurrently reducing the cooling effect from visible scattering.
      \item Thermal scattering decreases the absolute value of the mixing flux. In the middle atmosphere, thermal scattering amplifies the fractional change in temperature $T_{\rm e}/T_{\rm ne}-1$       caused by mixing flux. However, in the lower atmosphere, it attenuates the greenhouse effect driven by mixing flux. The combined impact of thermal scattering and vertical mixing within the lower atmosphere is less than the sum of their individual contributions, analogous to $1+1<2$ in the context of the greenhouse effect.  
   \end{enumerate}
The interaction between mixing flux and scattering is nonlinear. Evaluating the precise influence of mixing flux in the atmosphere requires simultaneous consideration of various physical processes. Additionally, in atmospheric retrieval, multiple processes, including mixing fluxes, should be analyzed concurrently to achieve a more accurate solution. 
In this study, we did not account for the effects of convection, as none of the atmospheric layers fulfilled the condition $\nabla > \nabla_{\rm ad}$ necessary for convection. Instead, aligning with the findings of \cite{Youdin2010} and \cite{2017Yu}, we observed the presence of a pseudo-adiabatic region in the lower atmosphere. The temperature gradient in this region closely follows the adiabatic gradient, albeit remaining slightly lower. 
In more general scenarios, there are many processes that need to be considered, such as horizontal transport processes brought about by atmospheric circulation, convective motions at the bottom of the atmosphere, etc. These fluxes also interact with both scattering and mixing fluxes. For example, vertical mixing can push the boundary of radiative convection into deeper atmospheric layers \citep{Youdin2010}. Future research will provide a comprehensive discussion of scenarios involving 
more processes.\par
Although $K_{\rm zz}$ is a spatially varying quantity \citep{2023Arfaux}, employing a constant value is adequate for qualitatively exploring the interplay between mixing processes and scattering. Variations in $K_{\rm zz}$ predominantly impact the strength of the findings. 
To simplify the calculation and align with the analytical results of \cite{Heng2014}, we adopt the semi-grey and Eddington approximations. However, it is important to note that these simplifications may impact our findings. For example, \cite{2015Parmentier} indicates that the relative uncertainty on the temperature profile resulting from the Eddington approximation is about several percent. Future work will focus on moving beyond these simplifications to develop a more realistic and self-consistent forward model. \par
\section*{acknowledge}
    We thank the anonymous referee for the helpful comments and suggestions that clarified and improved the manuscript.
    This work is supported by the National Natural Science Foundation of China (NSFC, No.12288102), the National SKA Program of China (grant No. 2022SKA0120101), the National Key R \& D Program of China (No. 2020YFC2201200), the science research grants from the China Manned Space Project (No. CMS-CSST-2021-B09, CMS-CSST-2021-B12, and CMS-CSST-2021-A10), International Centre of Supernovae, Yunnan Key Laboratory (No.202302AN360001), Guangdong Basic and Applied Basic Research Foundation (grant 2023A1515110805), and the grants from The Macau Science and Technology Development Fund, and opening fund of State Key Laboratory of Lunar and Planetary Sciences (Macau University of Science and Technology) (Macau FDCT Grant No. SKL-LPS(MUST)-2021-2023). C.Y. has been supported by the National Natural Science Foundation of China (grants 11521303, 11733010, 11873103, and 12373071).  

%






\appendix
\section{Linearization}\label{Linearization}
Equations (\ref{eq.RTE}, \ref{nbalance1}, \ref{nbalance2}) and their auxiliary expressions serve as the primary focus of the numerical calculations.
We linearize the equations and employ the Rybicki’s method \citep{1971Rybicki} following  \citet{2001Peraiah,Gandhi2017, Hubeny2017}. 
These involve discretizing the equations along the direction of increasing pressure, substituting differences for derivatives and sums for integrals, thereby leading to a series of nonlinear algebraic equations. The values of $J_{\rm \nu}$, and $T$ at the center of each atmospheric layer are yet to be determined. Additionally, the calculation of the mixing flux $F_{\rm eddy}$ requires evaluating the logarithmic temperature gradient $\nabla$ at the boundaries of each layer. The solution to these nonlinear algebraic equations is obtained using the Newton-Raphson method. Upon linearization, we derive a set of matrix equations for $J_{\rm \nu}$, $T$, and $\nabla$:
\begin{equation}\label{dec1}
    \mathbf{U}_{\rm k}\delta \mathbf{J}_{\rm k} + \mathbf{V}_{\rm k}\delta \mathbf{T} = \mathbf{E}_{\rm k} \ ,
\end{equation}
\begin{equation}\label{dec2}
    \sum _{\rm k}^2 \mathbf{X}_{\rm k}\delta \mathbf{J}_{\rm k} + \mathbf{A}\delta \mathbf{T} + \mathbf{W}\delta \mathbf{\nabla} = \mathbf{F} \ ,
\end{equation}
\begin{equation}\label{fnabla}
    \mathbf{Ad}\delta \mathbf{T} + \mathbf{Wd}\delta \mathbf{\nabla} = \mathbf{Fd} \ .
\end{equation}
The first matrix equation arises from linearizing the second-order Equation \eqref{eq.RTE}. 
The second matrix equation is derived from linearizing the radiative equilibrium conditions, Equations~\eqref{nbalance1} and \eqref{nbalance2}. The third matrix equation results from the definition of $\nabla$, Equation~\eqref{nabla}.

Linearized equation \eqref{eq.RTE}, we get
\begin{equation}
   \frac{f_{i-1,k}}{\Delta \tau_{i-1/2,k}\Delta \tau_{i,k}}J_{i-1,k}-\frac{f_{i,k}}{\Delta \tau_{i,k}}\left(\frac{1}{\Delta \tau_{i-1/2,k}}+\frac{1}{\Delta \tau_{i+1/2,k}}\right)J_{i,k}+ \frac{f_{i+1,k}}{\Delta \tau_{i+1/2,k}\Delta \tau_{i,k}}J_{i+1,k}-\gamma_{i,k}(1-\omega_{i,k})(J_{i,k}-B_{i,k})=E_{i,k} 
\end{equation}
 for layers: $0 < i < M$. For the boundaries $i=0$ and $i=MD$, Linearized equation \eqref{nbalance1} and \eqref{nbalance2}, we get
\begin{equation}
    \frac{f_{0,k}J_{1,k}-f_{1,k}J_{1,k}}{\Delta\tau_{1/2,k}}-\frac{\Delta\tau_{1/2,k}}{2}\gamma_{0,k}(1-\omega_{0,k})(J_{0,k}-B_{0,k})-\gamma_{0,k}(\mathrm{g}_{0,k}J_{0,k}-H_{k}^{\rm ext})=E_{0,k},
\end{equation}
and
\begin{align}
    \frac{f_{M-1,k}J_{M-1,k}-f_{M,k}J_{M,k}}{\Delta\tau_{M-1/2,k}}+\frac{\Delta\tau_{M-1/2,k}}{2}\gamma_{M,k}(1-\omega_{M,k})(J_{M,k}-B_{M,k})&\\ -
    \gamma_{M,k}\left[\frac{1}{2}(B_{M,k}-J_{M,k})+\frac{B_{M-1,k}-B_{M,k}}{3\Delta\tau_{M-1/2,k}}\right]&=E_{M,k}.
\end{align}
$\mathbf{E}_{\rm k}$ are the discrete residuals and they equal $0$ for the solutions that fit Equations (\ref{eq.RTE}, \ref{nbalance1}, \ref{nbalance2}). $\mathbf{X}_{\rm k}$ and $\mathbf{A}$ are variations of $\mathbf{F}$ with respect to $\mathbf{J}_{\rm k}$ and $\mathbf{T}$, respectively.
Similarly, $\mathbf{F}$ is the discrete residual when we discretized the energy conservation equations (Equation \eqref{nbalance1} for the upper atmosphere, and Equation \eqref{nbalance2} for the lower atmosphere). $\mathbf{X}_{\rm k}$, $\mathbf{A}$ and $\mathbf{W}$ are variations of $\mathbf{F}$ with respect to $\mathbf{J}_{\rm k}$, $\mathbf{T}$, and $\nabla$ respectively. $\mathbf{Fd}$ is the discrete residual when we discretized equation Equation~\eqref{nabla} and $\mathbf{Ad}$ and $\mathbf{Wd}$ are variations of $\mathbf{Fd}$ with respect to $\mathbf{T}$ and $\mathbf{\nabla}$, respectively.
The subscript $k$ denotes a specific frequency $\nu=\nu_{\rm k}$. Within the semi-grey approximation, $k$ only takes on two values ($k=1,2$), corresponding to frequency bands (``v'' and ``th''). The vectors $\mathbf{J}_{\rm k}= (J_{\rm 1,k}, J_{\rm 2,k},..., J_{\rm MD,k})$ include the mean intensity of frequency $k$ across each 
discretized layers of the atmosphere.\par
From Equation~\eqref{dec1} the vector $\delta \mathbf{J}_{\rm k}$ can be written as
\begin{equation}\label{J}
    \delta \mathbf{J}_{\rm k}=\mathbf{U}_{\rm k}^{-1}\mathbf{E}_{\rm k}-\mathbf{U}_{\rm k}^{-1}\mathbf{V}_{\rm k}\delta \mathbf{T} \ .
\end{equation}
Substituting Equation~\eqref{J} into Equation~\eqref{dec2}, one obtains
\begin{equation}\label{T}
    \left(\mathbf{A}- \sum _{\rm k}^2 \mathbf{X}_{\rm k}(\mathbf{U}_{\rm k}^{-1}\mathbf{V}_{\rm k})\right)\delta \mathbf{T}+\mathbf{W}\delta \mathbf{\nabla} =\mathbf{F}- \sum _{\rm k}^2 \mathbf{X}_{\rm k}(\mathbf{U}_{\rm k}^{-1}\mathbf{E}_{\rm k}) \ .
\end{equation}
Solving Equations \eqref{fnabla} and \eqref{T} yields the correction for the logarithmic temperature gradient and the temperature of each layer, denoted by $\delta \mathbf{\nabla}$ and $\delta \mathbf{T}$. $\delta \mathbf{J}_{\rm k}$ is determined from Equation \eqref{J}. This approach allows for efficient computation even when considering a large number of wavelengths, despite only two frequencies being utilized in this paper. This method is implemented to facilitate future expansions of the code. The updated temperature, mean intensity, and temperature gradient are obtained by applying the corrections $\delta \mathbf{T}$, $\delta\mathbf{J}_{\rm k}$, and $\delta \mathbf{\nabla}$ to $\mathbf{T}$, $\mathbf{J}$ and $\mathbf{\nabla}$ respectively. We iterate this process using the updated values until the temperature correction, $\delta \mathbf{T}/\mathbf{T}$, falls below a specified tolerance threshold.

\par


\bibliography{sample631}{}
\bibliographystyle{aasjournal}

\end{document}